\documentclass[twocolumn]{revtex4-2}

\usepackage{graphicx}
\usepackage{caption}
\usepackage{ragged2e}
\DeclareCaptionJustification{justified}{\justifying}

\begin{document}

\title{High-power, low-phase-noise, frequency-agile laser system for delivering fiber-noise-cancelled pulses for Strontium clock atom interferometry}

\author{Kenneth DeRose}
\author{Tejas Deshpande}
\author{Yiping Wang}
\author{Tim Kovachy}

\affiliation{The Center for Fundamental Physics and Department of Physics and Astronomy, Northwestern University, Evanston, IL 60208}

\begin{abstract}
We present the development of a laser system for performing single-photon atom interferometry on the 698\,nm clock transition in ultracold Strontium. We coherently combine the power of two Titanium:Sapphire lasers and demonstrate chirps of 200\,MHz in 2.5\,ms while phase-locked to an optical reference. Moreover, we demonstrate a novel scheme to deliver 4\,W \emph{pulsed} beams to the atoms via a mode-cleaning optical fiber using active noise cancellation.
\end{abstract}

\maketitle

\begin{figure*}[!htb]
		\begin{centering}
    \includegraphics[width=\textwidth]{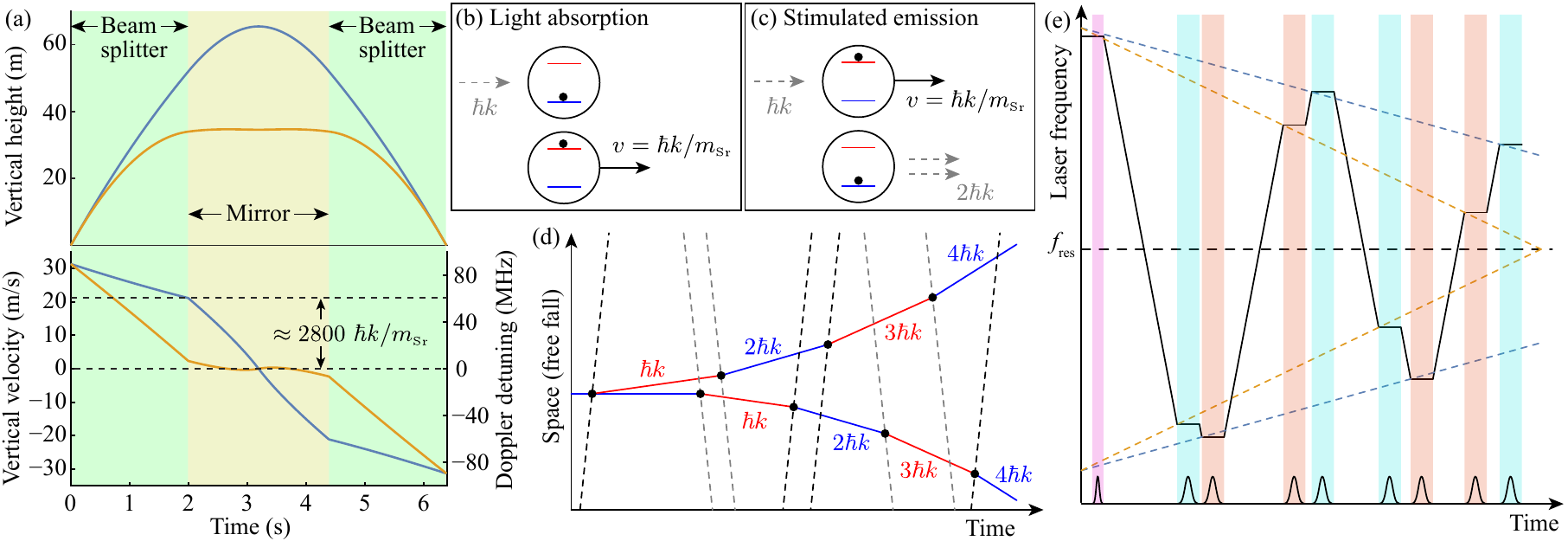}
		\end{centering}
    \caption{\textbf{(a)} Example AI sequence illustrating ultracold ${}^{87}{\rm Sr}$ atoms being split (BS sequence) over $\approx 30\,{\rm m}$, redirected towards each other (mirror sequence), and recombined (BS sequence) to generate an atomic interference pattern. The plots show simulated height and velocities (lab frame) of the two arms of the AI for MAGIS-100 with $f_{\rm Rabi} = 2\,{\rm kHz}$ and $\approx 2800\hbar k$ momentum splitting between interferometer arms (corresponding to in excess of $10^4$ total pulses), and $80\,{\rm MHz}/{\rm ms}$ laser chirp rate. $\hbar$ is the reduced Planck's constant and $k \approx 2\pi/(698\,{\rm nm})$. \textbf{(b) \& (c)} Cartoons of the atom-photon interaction occurring at black vertices in (d). Photons (gray dashed arrows) exchange momentum $\hbar k$ with atoms (circles) of mass $m_{\rm Sr}$, accompanied by the electron (black dot) transitioning between the excited (red) and ground (blue) states, such that said atom gets a velocity kick $v$. \textbf{(d)} Space time diagram in the freely-falling frame. Upward (downward) traveling photons are denoted by a dashed black (gray) line. The momenta ($n\hbar k$ with $n \ge 1$) and internal states (red/blue) of the atoms are indicated. Finite pulse width is ignored. \textbf{(e)} Schematic illustration of the evolution of the laser frequency for the AI sequence in (d). The \emph{qualitative} time evolution of the target Doppler shift (or velocity) is shown by (color-coded) dashed lines for the BS sequence from \textbf{(a)}.}
    \label{fig:atomic_physics}
\end{figure*}

% Physics introduction
Precision tests of fundamental physics place the highest demands on the properties of the utilized laser(s) such as high power, low phase noise, large frequency agility, and low-aberration spatial mode, not only to increase their metrological sensitivity, but also to suppress systematic errors \cite{Safronova_2018_RMP}. In particular, in a novel class of gravitational wave (GW) and dark matter (DM) \cite{Graham_2013_PRL,Arvanitaki_2018_PRD,MIGA_2018_SR,MAGIS_2021_QST,AION_2020_JCAP,AEDGE_2020_EPJ,ZAIGA_2020_IJMPD} detectors, lasers with the aforementioned properties are used to manipulate ultracold atomic waves. This technique, known as light-pulse atom interferometry (AI), uses laser pulses (``atom optics'') to act as effective beam splitters (BSs) and mirrors (Fig.\,\ref{fig:atomic_physics}) for the atomic wave packets freely falling under gravity \cite{tino_kasevich_book}. Lasers with these properties are also valuable for quantum sensors with practical applications \cite{Bongs_2019_NRP,Narducci_2022_APX,Wang_2022_RSI}.

% Atom interferometry details
Recently, atom interferometers using the ultra-narrow $698\,{\rm nm}$ clock transition of Strontium have emerged as a promising tool for applications including GW detection, DM detection, and foundational studies of quantum science \cite{Hu_2017_PRL,Hu_2019_CQG,MAGIS_2021_QST,AION_2020_JCAP,AEDGE_2020_EPJ,ZAIGA_2020_IJMPD}. These clock atom interferometers have demanding laser system requirements. Hundreds to thousands of laser pulses will be used to enhance the interferometer sensitivity \cite{Graham_2013_PRL,Arvanitaki_2018_PRD}. While differential measurement configurations can suppress the influence of laser phase noise on the atom interferometer phase \cite{Graham_2013_PRL,Arvanitaki_2018_PRD}, low laser phase noise remains important to minimize pulse transfer inefficiencies \cite{Chen_2012_PRA,Chiarotti_2022_PRXQ}. Moreover, frequency agility of around $100\,{\rm MHz}$ on millisecond timescales is required to account for the large Doppler shifts associated with the atoms' ballistic motion in tall atomic fountains and the velocity recoils received from the atom optics pulses (see Fig.\,\ref{fig:atomic_physics}). Furthermore, multi-Watt powers are needed to drive the weak clock transition.

\begin{figure*}[!htb]
		\begin{centering}
    \includegraphics[width=\textwidth]{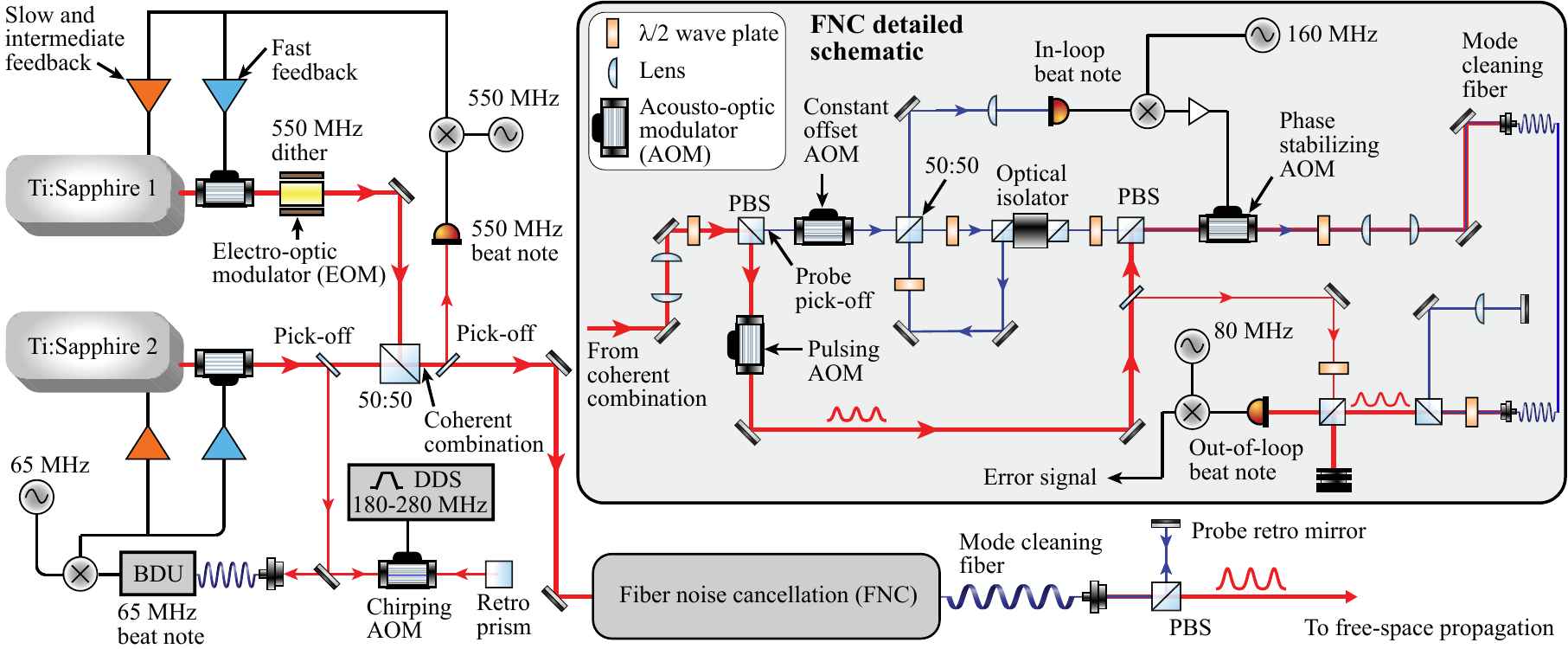}
		\end{centering}
    \caption{\textbf{Main schematic}: Optical layout for CC, offset-locking, and chirping. The BDU is configured for locking to a fixed 65 MHz beat note. \textbf{Inset}: detailed view of the FNC subsystem. The thickness of the lines representing optical paths is used to distinguish high (thick) and low (thin) power beams rather than being an indicator beam waist.}
    \label{fig:optics}
\end{figure*}

% Survey of lasers
Simultaneously achieving low phase noise and frequency agility for high-power lasers is challenging \cite{Wei_2015_OE}. Per the requirements outlined above, there are examples of lasers with: (a) low phase noise but low frequency agility \cite{Jiang_2010_OE}, (b) high frequency agility but comparatively high phase noise \cite{Theron_2015_APB,Wang_2022_RSI}, and (c) large frequency agility and low phase noise but low power \cite{Rohde_2014_OL,Menoret_2011_OL}. In this letter, we present the design of a high-power, frequency-agile, and low-phase-noise laser system.

% Fiber and solid-state lasers
Multi-Watt power requirements in the relevant wavelength range favor the use of solid-state or fiber lasers/amplifiers. The latter require a master oscillator power amplifier (MOPA) configuration \cite{Sane_2012_OE,Chiow_2012_OL,Kim_2020_OL}. However, the impact of noise properties of the fiber amplifiers in a given application needs to be carefully evaluated \cite{Cheng_2021_OE}. Moreover, due to the operating range of fiber amplifiers being in the near infrared, nonlinear frequency conversion is required.

% Advantages of Ti:Sa
Alternatively, one of highest performance solid-state lasers is Titanium:Sapphire (Ti:Sa). Not only do Ti:Sa operate in the visible spectrum, thus eliminating the need for nonlinear frequency conversion, but they also have a tuning range in the excess of $100\,{\rm nm}$. The latter is favorable for addressing different transitions (e.g. $679\,{\rm nm}$ and $689\,{\rm nm}$) in Strontium for DM detection \cite{MAGIS_2021_QST}. Furthermore, the robustness of the Ti:Sa crystal obviates the need for MOPA systems. Therefore, we decided to employ Ti:Sa technology for the MAGIS-100 atom optics laser system \cite{MAGIS_2021_QST}.

% Case against intra-cavity EOM
To simultaneously achieve our active phase noise reduction and chirping requirements, our Ti:Sa system implements combined feed forward (FF) and feedback using piezoelectric transducers (PZTs) to achieve a rapidly tunable offset lock from a stable optical reference, circumventing the limited frequency tuning range of acousto-optics modulators (AOMs) \cite{Wang_2022_RSI} and the limited power handling capabilities of fiber electro-optic modulators (EOMs). For narrow-range, high-bandwidth feedback, we used external acousto-optic modulators AOMs, trading the higher bandwidth of intra-cavity EOMs \cite{Muller_2006_OL} for power.

\begin{figure*}[ht]
		\begin{centering}
    \includegraphics[width=\textwidth]{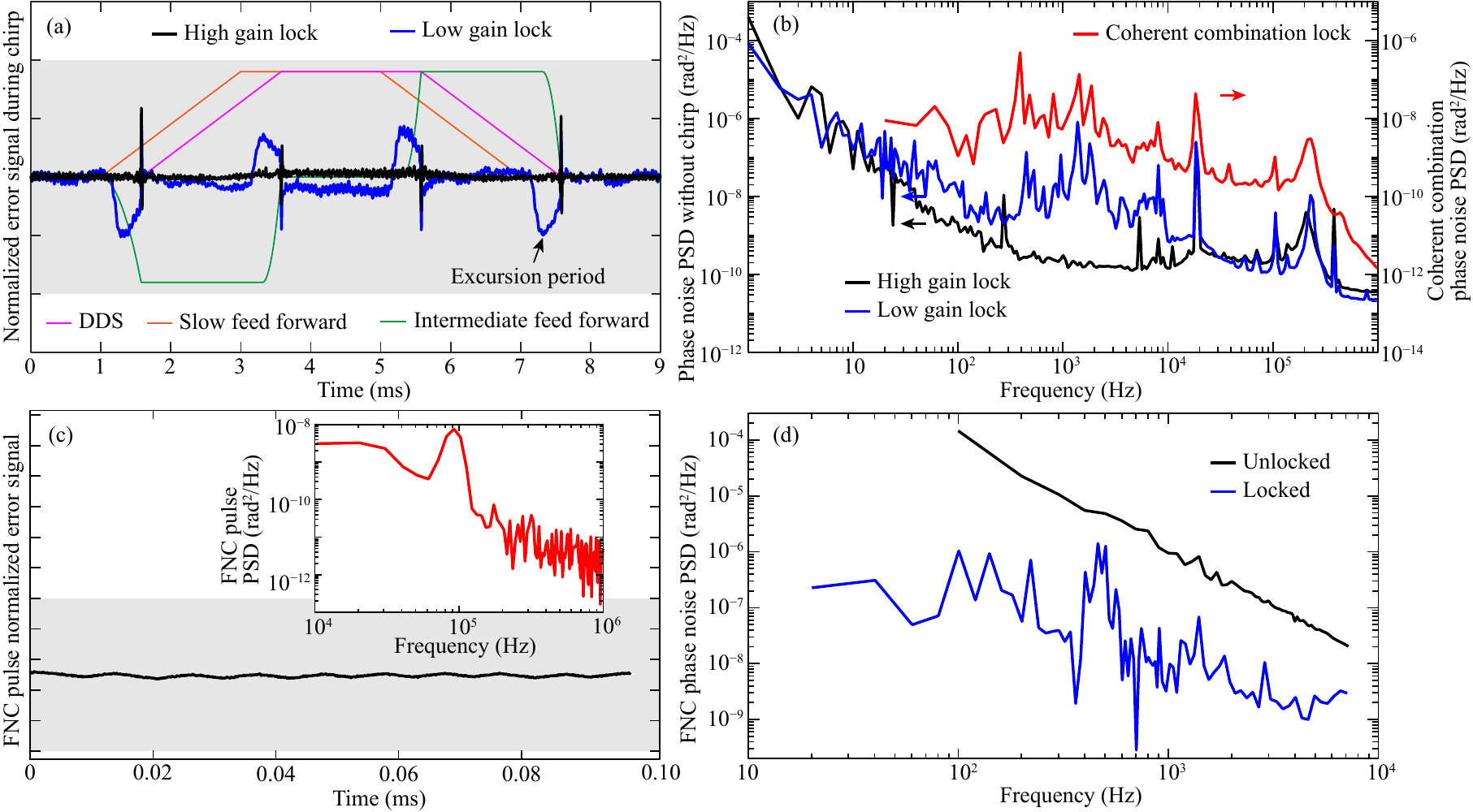}
		\end{centering}
    \caption{\textbf{(a)} Error signals in HGM \& LGM during sequential upward and downward frequency chirps (see main text for details), with the DDS and FF signals' temporal profiles overlaid as guides to the eye. The \textbf{(b)} PSD of the offset locks (HGM \& LGM) and \& CC. The $y$-axis of the latter is vertically shifted for clarity. The peak at 19 kHz arises due to internal etalon dither of Ti:Sa. \textbf{(c)} FNC lock error signal over a single $100\,\mu{\rm s}$ pulse. Inset shows the PSD of the single-pulse error signal. \textbf{(d)} PSDs of the locked and unlocked error signals for a train of $0.1\,{\rm ms}$ pulses with an $8\,{\rm kHz}$ repetition rate. Data points during the inactive pulse times were removed and a non-uniform fast Fourier transform algorithm was performed on the remaining data set to generate the PSDs. The shaded regions in (a) and (c) represent the full phase range. All phase noise measurements are made out-of-loop.}
    \label{fig:data}
\end{figure*}

% Motivate fiber noise cancellation
Moreover, in many applications, a low-distortion spatial mode is important for reducing systematic errors. This is often achieved via spatially filtering the beam with an optical fiber. In applications such as AI that require laser pulses, it is favorable to place the pulse actuator (e.g., AOM) before the fiber to minimize post-fiber optical elements that can add spatial distortions \cite{Kovachy_2015_nature}. However, fiber transmission adds significant phase noise to the laser beam from environmental temperature and acoustic perturbations. Therefore, active fiber noise cancellation (FNC) is required to maintain the upstream low phase noise properties of the beam. Previous FNC methods were designed for constant power laser transmission through the fiber \cite{Ma_1994_OL}. Here, we present a new FNC method suitable for the case of pulsed beams.

% Introduction to optical layout
Fig.\,\ref{fig:optics} shows a schematic representation of our laser system. It consists of three primary subsystems: (1) coherent combination (CC) of optical power, (2) offset locking to and chirping with respect to an optical reference, and (3) noise cancellation through a mode-cleaning fiber. We use commercial single-frequency continuous wave Ti:Sa lasers from \emph{M Squared Lasers}. Using an $18\,{\rm W}$ optical pump at $532\,{\rm nm}$, produced by a diode-pumped solid state laser (Equinox), the Ti:Sa bow tie cavity (SolsTiS) generates $\approx 5\,{\rm W}$ of optical power at $698\,{\rm nm}$; this wavelength approximately corresponds to the resonance frequency, $f_{\rm res}$, (Fig.\,\ref{fig:atomic_physics}\,(d)) of the clock transition in ${}^{87}{\rm Sr}$. The SolsTiS resonator length is stabilized by a stack of two PZTs. These PZTs provide feedback on slow and intermediate time scales; fast feedback is provided by external AOMs. The low and high bandwidth PZTs have a tuning range of $25\,{\rm GHz}$ and $80\,{\rm MHz}$ respectively.

We boost the power by combining the output of the two Ti:Sa lasers \cite{Ma_2011_OL}. Ti:Sa 1 is phase-locked to Ti:Sa 2 such that the powers of both lasers is coherently combined on the 50:50 BS with a 90\% efficiency. The CC error signal is derived from a 550 MHz EOM dither. The dither frequency is chosen to be far-detuned from all Doppler-detunings of the atomic transition [see Fig. \ref{fig:atomic_physics}(a)]. An optimal lock produces $7\,{\rm W}$ of coherently combined power with (downstream) $3.9\,{\rm W}$ incident on the 50:50 BS from each Ti:Sa. The measured power spectral density (PSD) of the phase noise is shown in Fig.\,\ref{fig:data}\,(b).

% Offset lock and chirping
The frequency of Ti:Sa 2 ($f_{{\rm L}2}$) is offset-locked to a tooth ($f_{{\rm OFC}}$) of a commercial OFC from \emph{Menlo Systems GmbH} (FC1500 Quantum, stabilized to an ORS-Cubic High Performance 1550 optical reference system), via the beat detection unit (BDU), using a 2\% pick-off from the main beam. Thus, due to the phase lock between the two Ti:Sa ($f_{{\rm L}1} = f_{{\rm L}2}$), the CC derives its stability from the OFC as well. In order to meet our chirping requirements, we double-pass the Ti:Sa 2 pick-off through an AOM driven by a frequency-agile ($f_{{\rm DDS}}$) direct digital synthesizer (DDS) before feeding it to the BDU. As a result, we get 
\begin{equation}
    f_{{\rm L}2}(t)=f_{{\rm OFC}}+2f_{{\rm DDS}}(t)+65\,{\rm MHz} \label{eq:chirp}
\end{equation}
where $t$ is time. The laser chirping profiles illustrated in Fig.\,\ref{fig:atomic_physics}\,(e) can be generated with a maximum ${\rm d}f_{{\rm L}2}(t)/{\rm d}t\approx 80\,{\rm MHz}/{\rm ms}$. This chirping scheme allows us to sweep the \emph{main} beam over $200\,{\rm MHz}$ without suffering diffraction loses from the chirping AOM.

Eq.\,(\ref{eq:chirp}) is physically implemented with three feedback loops: slow, intermediate, and fast. The \emph{M Squared Lasers} Phase Lock ICE BLOC servo controller \cite{PID_book} can operate in two modes: high \& low gain mode (HGM \& LGM). Fig.\,\ref{fig:data}\,(b) shows the PSD of the offset lock of Ti:Sa 1 to Ti:Sa 2 in both modes; the latter is a proxy for the OFC. Note the 2 orders of magnitude lower PSD in the HGM in the kHz range.

We implement a FF scheme to compensate for the limited bandwidth of the dual-PZT stack in the SolsTiS during a chirp. Fig.\,\ref{fig:data}\,(a) shows the profile of the FF signals applied to the two PZTs, for a nominal $f_{{\rm DDS}}(t)$ profile qualitatively similar to Fig.\,\ref{fig:atomic_physics}\,(e). The resulting error signals are shown in HGM and LGM. The HGM lock curve corresponds to frequency chirps over a 140\,MHz range, while the LGM lock curve corresponds to frequency chirps over a 200\,MHz range. In the low gain lock trace, four phase excursions spanning 0.5\,ms are observed, caused by imperfections in the FF ramps. At the end of each excursion, there is a transient in phase due to the DDS suddenly starting or ending the chirp. No pulses are sent to the atoms during the low gain lock excursion period or phase spikes. In LGM, an effective chirp rate (accounting for the 0.5\,ms phase excursions) of 200\,MHz in 2.5\,ms (80\,MHz/ms) is achieved. In HGM, the increased gain suppresses the phase excursions but also prevents the lock from holding reliably over a full 200\,MHz sweep in 2\,ms. The sweep rate is therefore reduced to 70\,MHz/ms in HGM so that the locking remains robust. The choice of HGM versus LGM will depend on the details of a particular application.

% FNC intro
The free-space chirped, offset-locked, and coherently combined beam is coupled into a short (high-power) fiber (Thorlabs PM630-HP) before being delivered (free-space) to the ${}^{87}{\rm Sr}$ atoms. Intermediate downstream fiber transmission ensures that the atoms see a low-aberration spatial mode. In order to preserve the pre-fiber low phase noise properties of the beam, said fiber transmission requires active noise cancellation of environmental thermal and acoustic perturbations \cite{Ma_1994_OL}. The inset of Fig.\,\ref{fig:optics} shows a novel FNC scheme we implement for transmitting high-power pulses required for AI sequences like those shown in Fig.\,\ref{fig:atomic_physics}\,(d).

% Probe beam
A unique feature of our FNC scheme is the use of a constant-power probe beam (false color blue, $2\%$ power relative to main beam) to enable the feedback to be in constant operation, independent of the pulsing of the main beam (false color red). This avoids phase transients associated with intermittent lock acquisition as the main beam pulses \cite{PID_book}. The retro-reflected probe beam, containing fiber-induced phase noise, is interfered with the original probe on a 50:50 BS, and the resulting error signal (derived from the in-loop beat note) is fed back to the phase stabilizing AOM, thus canceling the noise in the main beam as well. The performance of this FNC scheme is evaluated by measuring an out-of-loop error signal generated by beating the main beam against picked off light that does not pass through the fiber. The time- and frequency-domain performance of the FNC subsystem is shown in Fig.\,\ref{fig:data}\,(c) and (d).

% Isolate probe and main beams
We implement the following features to prevent the main and probe beams from influencing each other. We use a polarizing BS (PBS) to make the polarizations of the probe and main beams orthogonal. Moreover, we use an AOM to maintain a constant frequency offset ($180\,{\rm MHz}$) between the two beams. This prevents the main beam from leaking into the probe's in-loop beat note signal.

% Power to atoms
The power transmission efficiency of the FNC subsystem is 57\% (two AOMs at 90\% efficiency and a fiber coupling efficiency of 70\%). Therefore, a $4\,{\rm W}$ coherently combined beam will be delivered to the ${}^{87}{\rm Sr}$ atoms.

% Conclusion
In summary, we have presented the design and performance of a Ti:Sa laser system with high power, low phase noise, large frequency agility, and low-aberration spatial mode for Strontium clock atom interferometry. This laser system will be deployed in the MAGIS-100 experiment \cite{MAGIS_2021_QST}. Even higher power could be achieved by coherently combining more than two lasers. Moreover, increased chirp rate can, in principle, be achieved by more sophisticated FF techniques involving stacks of more than 2 PZTs.

% Acknowledgments and other info
\hfill

\noindent\textbf{Funding.} This work was supported by Gordon and Betty Moore Foundation Grant GBMF7945 and a David and Lucile Packard Fellowship for Science and Engineering.

\hfill

\noindent\textbf{Acknowledgments.} We thank Jason Hogan and M Squared Lasers (in particular Joseph Thom) for valuable discussions and technical assistance.

\hfill

\noindent\textbf{Disclosures.} The authors declare no conflicts of interest.

\hfill

\noindent\textbf{Data availability.} Data underlying the results presented in this paper can be obtained upon reasonable request from the authors.

\bibliography{magis_698_nm}

\end{document}